\documentclass[12pt]{article}
\usepackage[T1]{fontenc}
\usepackage{geometry}
\usepackage{amsmath}
\makeatletter

\providecommand{\LyX}{L\kern-.1667em\lower.25em\hbox{Y}\kern-.125emX\@}
\newcommand{\noun}[1]{\textsc{#1}}

\newcommand{\lyxaddress}[1]{
  \par {\raggedright #1 
  \vspace{1.4em}
  \noindent\par}
}

\makeatother

\begin{document}

\title{Geometric quantisation of the global Liouville mechanics}

\author{Z. Bajnok, D. Nógrádi, D. Varga, F. Wágner}

\maketitle

\lyxaddress{\emph{Institute for Theoretical Physics, Roland Eötvös University, H-1117 Budapest,
Pázmány sétány 1/A, Hungary }}

\def\R{{\bf R}} 

\def\Tr{{\rm Tr}} 

\def\sl2{$SL(2,{\bf R})$}

\def\AP#1#2#3{{\rm Ann. of Phys.\ }{\bf #1} (#2) #3}  

\def\JMP#1#2#3{{\rm Jour. of Math. Phys.\ }{\bf #1} (#2) #3} 

\begin{abstract}
The reduced \noun{\sl2} WZW quantum mechanics is analysed within the framework
of geometric quantization. The spectrum of the Hamiltonian is determined, and
it is found, that in contrast to previous approaches, there is a unique, physically
preferred quantisation of the system. 
\end{abstract}
The global aspects of the WZW \( \rightarrow  \)Toda reduction, \cite{BaFeFo},
were analysed in \cite{FeTs} at the classical level. As a toy example the authors
also considered the reduction of the system of a free particle moving on the
\sl2 group manifold. The problem of how to quantize this reduced system was
addressed in \cite{KoTs,Tam}. Both found inequivalent quantisations labelled
by one or two real parameters. In this letter, using the method of geometric
quantisation, we argue that there is a unique quantisation of the problem. 

After reminding the known facts about the dynamics of a free particle moving
on the \sl2 group we impose the Liouville type constraints. The reduced system
is quantised geometrically and the spectrum of the Hamiltonian is determined.
Finally we compare the results obtained to those of \cite{KoTs,Tam} and comment
on the differences. 

Consider a free particle moving on the group \sl2. The dynamics is governed
by the Lagrangian:

\[
L(g,\dot{g})=\frac{1}{2}\Tr (g^{-1}\dot{g}g^{-1}\dot{g}).\]
It is invariant under the transformations \( g\rightarrow g_{left}gg_{right}^{-1} \),
for which the Noether currents are \( J:=J_{left}=\dot{g}g^{-1} \) and \( \tilde{J}:=J_{right}=-g^{-1}\dot{g}=-g^{-1}Jg \).
The equation of motion is the conservation of the current: \( \dot{J}=0 \),
with the geodetical motions as solutions. 

In describing the dynamics at the Hamiltonian level we use the velocity phase
space \( M=TG=G\times \cal G \) coordinatised by \( g \) and \( J\in \cal G \)
. The symplectic form is \( \Omega =d\theta _{L}=d\Tr (Jdgg^{-1}) \) as in
\cite{HST}, where \( \theta _{L}=\sum _{i}\frac{\partial L}{\partial q_{i}}dq_{i} \)
is the symplectic potential determined by the Lagrangian as usual. Clearly the
symplectic potential is not unique. Due to the nontrivial topology of the phase
space \( M=(S^{1}\times \R ^{2})\times \R ^{3} \) we could have different choices
for \( \theta  \) corresponding to the various cohomology elements labelled
by \( \R  \) . Note however, that the only symplectic potential which is invariant
under the full symmetry group is the one that corresponds to the Lagrangian,
i.e. \( \theta _{L} \). (To see that no other invariant symplectic potential
exists observe, that thanks to the structure of the phase space the nontrivial
elements of the cohomology group correspond to the nontrivial cohomology class
of the group itself. On the group however there exist no nontrivial left and
right invariant one-form). The Hamiltonian \( H=\frac{1}{2}\Tr (J^{2}) \) generates
the dynamics: \( \dot{J}=0, \) \( \dot{g}=Jg \) . 

The global Liouville system can be obtained by Hamiltonian reduction. One fixes
certain components of the conserved currents:
\[
J=\left( \begin{array}{cc}
J_{0} & J_{+}\\
1 & -J_{0}
\end{array}\right) \quad ;\quad \tilde{J}=\left( \begin{array}{cc}
\tilde{J}_{0} & 1\\
\tilde{J}_{-} & -\tilde{J}_{0}
\end{array}\right) .\]
The projected symplectic form is degenerate, the constraints are first class.
Since the gauge transformations are nothing but the symmetry transformations
generated by the strictly upper and strictly lower diagonal matrices the Hamiltonian
and the symplectic potential are gauge invariant. The reduced system can be
obtained via gauge fixing: 
\[
J_{gf}=\left( \begin{array}{cc}
0 & J_{+}\\
1 & 0
\end{array}\right) \quad ;\quad \tilde{J}_{gf}=\left( \begin{array}{cc}
0 & 1\\
J_{+} & 0
\end{array}\right) \quad ;\quad g=\left( \begin{array}{cc}
-J_{+}g_{22} & -g_{21}\\
g_{21} & g_{22}
\end{array}\right) .\]
The parameters \( g_{21},g_{22},J_{+} \) are not independent, we have the condition:
\begin{equation}
\label{det}
g^{2}_{21}-J_{+}g_{22}^{2}=1.
\end{equation}
These mean that the reduced phase space is a regular hypersurface of \( \R ^{3} \)
determined by (\ref{det}). Topologically it is \( S^{1}\times \R  \) and has
the symplectic potential:

\begin{equation}
\label{pot}
\theta =-2J_{+}g_{22}dg_{21}+g_{21}g_{22}dJ_{+}+2J_{+}g_{21}dg_{22}.
\end{equation}
Let us emphasize once more that it is possible to change the cohomology class
of \( \theta  \) without affecting the symplectic structure. This altered symplectic
potential however, does not correspond to the original \( \theta _{L} \), i.e.
to the Lagrangian, so it describes a physically different system as we will
see at the quantum level. 

It was shown in \cite{RaYa}, that the reduced phase space is in fact symplectomorphic
to \( T^{*}S^{1} \), but the Hamiltonian is very complicated: \( H=\sin ^{-2}\varphi (\cos ^{2}\varphi -\exp (p_{\varphi }\sin ^{2}\varphi )) \),
so this description is not useful in quantising the system. The Hamiltonian
in our language however is very simple:
\[
H=\frac{1}{2}\Tr (J_{gf}^{2})=J_{+}.\]

The solutions of the equations of motion can be read off from eq. (\ref{det}),
they are the curves with constant energies. They are ellipsis for negative energies,
lines for zero energy and hyperbolas for positive energies. We adopt a coordinate
system to respect these curves, that is, one of the parameters is the energy,
while the other one parametrizes the orbit. The phase space can be covered by
four neighbourhoods as:
\[
M_{<}^{\pm }=\{H<0,\quad \pm g_{21}>-\epsilon \}\quad ;\qquad M_{>}^{\pm }=\{H>-\epsilon ,\quad \pm g_{21}>1-\epsilon \},\]
where \( \epsilon  \) is a small positive parameter less than \( 1/2 \). From
now on in the expression \( \pm  \), \( + \), (\( - \)) will refer to the
part where \( g_{21}>0 \), (\( <0 \)), respectively. For negative energies
we parametrize the phase space as: 
\[
g_{21}=\pm \cos (\sqrt{-H}t)\quad ,\quad g_{22}=\sqrt{-H}^{-1}\sin (\sqrt{-H}t)\quad ;\qquad -\frac{\pi }{2}-\delta (\epsilon )<\sqrt{-H}t<\frac{\pi }{2}+\delta (\epsilon ),\]
while for positive energies as: 
\[
g_{21}=\pm \cosh (\sqrt{H}t)\quad \, \quad g_{22}=\pm \sqrt{H}^{-1}\sinh (\sqrt{H}t)\quad ;\qquad -\infty <t<\infty .\]
The symplectic potential and the symplectic form in both regimes are: 
\[
\theta =2Hdt+tdH\quad ;\qquad \Omega =dH\wedge dt.\]
Although this parametrization covers the whole phase space with a smooth limit
for \( H\rightarrow 0 \) the correct description is to restrict them into the
neighbourhoods defined above. The Hamiltonian vector field of the Hamiltonian,
\( H \), is defined in the usual way, \( \Omega (X_{H},\cdot )+dH=0 \) , it
is simply \( X_{H}=\frac{\partial }{\partial t} \). 

In quantising the theory we use the method of geometric quantization \cite{Wh,'S}.
It contains two steps. In the first, called the prequantization, one constructs
a hermitian line bundle over the phase space with curvature \( \hbar ^{-1}\Omega  \),
(with connection one-form \( \hbar ^{-1}\theta  \)). For each classical observable,
\( f \), a symmetric operator, \( \hat{f}=-i\hbar \nabla _{X_{f}}+f \), is
assigned, which acts on the square integrable sections of the bundle (pre-quantum
wave functions). This operator is self-adjoint if the Hamiltonian flow of \( f \)
is complete.  Pre-quantum wave functions, however depend on both coordinates,
which is not acceptable from an adequate quantum theory. To avoid this in the
second step a polarization is chosen and the Hilbert space is restricted to
be built up from the square integrable polarized sections. Accordingly only
those functions are quantisable, whose Hamiltonian flow preserves the polarization. 

The first step in quantizing the theory is easy. Since the symplectic form is
exact the line bundle exists and is also trivial. It is not unique however,
the inequivalent choices are characterised by the holonomy of the connection
and are parametrized by the unit circle. 

Similar situation appears in the case of non simply-connected configuration
space, like in the Bohm-Aharonov effect. There one is faced to the fact that
although at the classical level the system is determined by the equation of
motion, which can be derived from various Lagrangians, at the quantum level
the Lagrangian itself defines the theory, since it contains the vector potential
explicitly. Different Lagrangians -- different vetorpotentials -- with the same
classical theories describe different, physically not equivalent quantum systems. 

Concretising to our case we have to use the globally defined symplectic potential
(\ref{pot}), which corresponds to the original Lagrangian. Chosen a global
section \( s_{0} \) the connection is defined by:
\[
\nabla s_{0}=-i\hbar ^{-1}\theta s_{0}\quad ;\qquad \nabla s=(d-i\hbar ^{-1}\theta )s.\]
Since \( \theta  \) is real \( s_{0} \) can be normalized as \( (s_{0,}s_{0})=1 \).
Now each section can be written in the form \( s=\psi s_{0} \) , where \( \psi  \)
is a function on the phase space, i.e. the Hilbert space of the prequantized
theory consists of the square integrable functions (for the integration measure
\( h^{-1}\Omega  \)). The operator we are interested in is the Hamiltonian.
It acts on the sections as:
\[
\hat{H}s=(-i\hbar \nabla _{X_{H}}+H)s.\]
In order to make correspondence with the usual language of quantum mechanics
we remind, that
\[
\nabla _{X_{H}}s=\nabla _{X_{H}}\psi s_{0}=(X_{H}\psi -i\hbar ^{-1}\theta (X_{H})\psi )s_{0}=(\frac{\partial \psi }{\partial t}-i\hbar ^{-1}2H\psi )s_{0}\]

The second part of the geometric quantisation is much more delicate. Firts of
all we have to choose a polarization, a Lagrangian integrable distribution,
for which the Hamiltonian is quantisible, i.e. which is preserved by the flow
of the Hamiltonian. The most natural choice corresponds to \( X_{H} \), that
is the polarization given by \( P=\{\frac{\partial }{\partial t}\} \). It is
not a reducible polarization in the sense of \cite{Wh}, since the space of
leaves is not a Hausdorff topological space. The leaves corresponding to the
free motions, \( H=0 \), do not have disjoint neighbourhoods, they are connected
via the negative energy part of the phase space. It is not a problem however,
since for negative energies the leaves are compact and, as we will see soon,
no nonzero smooth covariantly constant section exists. 

The polarized sections, \( \nabla _{X_{H}}s=0 \) , have the form \( e^{2iH\hbar ^{-1}t}\psi (H)s_{0} \)
. The pairing of two polarized sections are constant along the leaves, (\( t=const. \)
surfaces), so if the leaves are not compact then the integral over them is infinite,
which is unacceptable. Clearly what we need is to integrate over the reduced
space \( M/P \) only, but there is no natural measure on this space. This problem
is solved in the framework of geometric quantization by changing the geometric
character of the wave functions: they are the square integrable polarized sections
of the modified bundle \( B\otimes \delta _{P} \) , where \( \delta _{P} \)
is the half-form bundle. This modification not only ensures that the pairing
of two wave functions gives densities on the space of leaves but also restore
the correct relationship between the quantizations corresponding to different
polarizations. 

Taking a look on the polarized sections we can see that for compact leaves covariantly
constant sections exist only for certain discrete values of the energy, consequently
they cannot be smooth. The proper description is to deal with distributional
wave functions in this domain, see \cite{'S} for the details. The supports
of these functions, the so called Bohr-Sommerfeld (BS) varieties, are determined
by the condition that the holonomy of the connection for the integral manifolds
of the polarization to be trivial:
\[
\oint _{\gamma }\theta =-2\sqrt{-H}2\pi =2\pi \hbar (n_{\gamma }+d_{\gamma })\quad ;\qquad n_{\gamma }\in Z,\]
where \( d_{\gamma }=\frac{1}{2} \) is the holonomy corresponding to the bundle
\( \delta _{P} \) as a consequence of the metaplectic correction. (In defining
the square root \( \delta _{P} \) of the canonical bundle \( K_{P} \) we have
to take into account that not the symplectic, rather its double cover, the metaplectic
group acts on it). For each BS variety a normalized section is associated, which
is the eigenfunction of the energy operator with the eigenvalue:
\[
H_{n}=-\frac{\hbar ^{2}}{4}(n+\frac{1}{2})^{2}\]
(This result can be understood also by introducing the new variables \( \omega =2\sqrt{-H},\varphi =\sqrt{-H}t \)
on the negative energy part of the phase space. Then it is easy to see that
this part of the phase space is the same as the phase space of the harmonic
oscillator parametrized by its energy \( \omega  \) and by the usual angle
\( \varphi  \) ).

The positive part of the spectrum is doubly degenerate and continuous. 

The whole Hilbert space can be built up from the energy eigenfunctions as follows.
It consists of the square integrable functions \( \Psi _{\pm }(H),\, H\geq 0 \)
and the square summable sequence of numbers \( \Psi _{n} \) . The inner product
of two elemenst of the Hilbert space, \( \Psi  \) and \( \Phi  \) is given
by:
\[
\langle \Psi ,\Phi \rangle =\int _{0}^{\infty }dH\Psi _{+}^{*}(H)\Phi _{+}(H)+\int _{0}^{\infty }dH\Psi _{-}^{*}(H)\Phi _{-}(H)+\sum _{n=0}^{\infty }\Psi _{n}^{*}\Phi _{n}.\]

Summarizing we quantized the system in the framework of geometric quantization
and found a unique discrete spectrum for negative energies and a continous doubly
degenerate spectrum for nonnegative energies. Comparing the result with that
of \cite{KoTs,Tam} we can say that our result is different from those: we have
a unique quantization. The appearance of their parameters is due to the fact
that they chose a bad parametrization for the phase space in which the lines
\( g_{22}=0 \) had to be removed. Since every motion intersects at least one
of those lines the flow of the Hamiltonian is not complete, that is the associated
operator is not self-adjoint. Finding the possible self-adjoint extensions they
arrived at the models, which in our language can be described by different symplectic
potentials. (The holonomy of the various symplectic potentials are parametrized
by a unit modulus number which appears also in the spectrum of the energy).
We have seen however that there is one physically preffered amoung them\c{ }
the one which corresponds to the Lagrangean. 

The authors thank to L. Fehér for the reading of the manuscript and for taking
our attantion for ref. \cite{RaYa}. Z. B. was supported by OTKA F019477,D25517,T029802.

\end{document}